\documentclass[twocolumn,10pt]{article} 
\usepackage{times}
\usepackage{vmargin}%
\usepackage[pdftex]{graphicx}

\usepackage{amsmath, amsthm, amssymb}
\usepackage[ruled, vlined]{algorithm2e}
\usepackage[pdftex,bookmarksnumbered=true]{hyperref}

\usepackage{float}
\floatstyle{ruled}
\newfloat{figure}{H}{fig}
\floatname{figure}{Figure}
\setmargrb{0.72in}{0in}{0.72in}{1in}

\date{}
\title{Multiprocessor Global Scheduling on Frame-Based DVFS Systems}
\author{Vandy \textsc{Berten}\footnote{Universit\'e Libre de Bruxelles, Fonds National de la Recherche Scientifique} \\\texttt{vandy.berten@ulb.ac.be} \and Jo\"el \textsc{Goossens}\footnote{Universit\'e Libre de Bruxelles}\\\texttt{joel.goossens@ulb.ac.be}  
}

\newcommand{\cpu}{\Pi}
\newcommand{\tasks}{\mathcal{T}}
\newcommand{\task}{\tau}
\newcommand{\textcpu}{\textsc{cpu}}

\begin{document}
\maketitle

\section{Introduction}

Nowadays, it is straightforward that energy efficiency is a crucial aspect of embedded systems where a huge number of small and very specialized autonomous devices interacting together through many kinds of media (wired/wireless network, bluetooth, GSM/GPRS, infrared\dots).
Moreover, we know that the uniprocessor paradigm will no longer hold in those devices. Even today, a lot of mobile phones are already equipped with several processors.

In this ongoing work, we are interested in multiprocessor energy efficient systems, where task durations are not known in advance, but are know stochastically. More precisely, we consider global scheduling algorithms for frame-based multiprocessor stochastic DVFS (Dynamic Voltage and Frequency Scaling) systems. Moreover, we consider processors with a \emph{discrete} set of available frequencies.

In the past few years, a lot of work has been provided in multiprocessor energy efficient systems. Most work was done considering static partitioning strategies, meaning that a task was assigned to a specific processor, and each instance of this task runs on the same processor. First of those work where devoted to deterministic tasks (with a task duration known beforehand, or the worst-case is considered), such as \cite{Aydin03, Yang05, Chen05, Chen05b}, and later probabilistic models were also considered \cite{Xian07, Mishra03}.
Only a little work has been provided about global scheduling, such as \cite{Chen04}, but for deterministic systems, or \cite{Dakai01}, using some slack reclamation mechanism, but not really using stochastic information.

As far as we know, no work has been provided with global scheduling on stochastic tasks. We propose to work towards this direction. Notice that the frame-based model we consider in our work, where every task share the same period, is also used by many researchers, such as \cite{Yang05, Chen04, Mishra03, Dakai01}. 

% \subsection{Related Work}
% *\cite{Aydin03}: Static partitionning of deterministic tasks, using slack reclamation when WCEC is not reach. They consider continuous frequencies between 0 and max. Offset=0.
% 
% *\cite{Xian07}: Static partitionning of probabilistic tasks. Discrete \textcpu{} frequencies.
% 
% *\cite{Yang05}: Static partitionning of deterministic tasks, on frame-based systems. Each \textcpu{} can be put on dormant mode, but all running \textcpu{} share the same frequency, which ranges continuously between $0$ and $\infty$.
% 
% \cite{Chen04}: They consider both static partitionning and some kind of global scheduling (static partitionning but with task migration), for a frame-based deterministic system. Continuous frequencies, first from $0$to $\infty$, then from $0$ to max.
% 
% *\cite{Mishra03}: Static partitionning (??), Frame-based, with precedence graph, using worst case and average execution time, continuous frequencies from $0$ to max. Mainly based on slack reclamation.
% 
% *\cite{Chen05}: Static partitionning, deterministic tasks, frequencies ranging from $0$ to $\infty$.
% 
% *\cite{Chen05b}: Extension of \cite{Chen05} where tasks have different power characteristics
% 
% \cite{Dakai01}: Global Scheduling and Static Paritionning. Frame-based, using slack reclamation, with/without precedence constraint.

\section{Model}
We consider $n$ sequential tasks $\task_1, \dots, \task_n$. Task $\task_i$ requires $x$ cycles with a probability $c_i(x)$, and its maximum number of cycles is $w_i$ (Worst Case Execution Cycles, or WCEC). The number of cycles a task requires is not known before the end of its execution. We consider a \emph{frame-based} model, where all tasks share the same deadline and period $D$ and are synchronous. In the following $D$ denote the frame length.

Those tasks run on $m$ identical \textcpu{} $\cpu_1, \dots, \cpu_m$, and each of those \textcpu{} can run at $M$ frequencies $f_1, \dots, f_M$. 

We consider that tasks cannot be preempted, but different instances of the same task can run on different processors, i.e., task migrations are allowed. We consider global scheduling techniques which schedule a queue of tasks ; each time a \textcpu{} is available, it picks up the first task in the queue, choose a frequency, and run the job. We assume the system is expedient\footnote{An expedient system is a system where tasks never wait intentionally. In other words, if a task is ready, the processor cannot be idle.}, and the job order has been chosen beforehand, but in some cases, in order to ensure the schedulability, the scheduler can adapt that order. In other words, we assume that the initial task order is not crucial and can be considered to be a soft constraint.

\section{Global Scheduling Algorithm}

In~\cite{RTCSA08}, we have provided techniques allowing to schedule such a task set on a \emph{single} \textcpu. The main idea is to compute (offline) a function giving, for each task, the frequency to run the task based on the time elapsed in the current frame. This function, $S_i(t)$ gave the frequency at which $\task_i$ should run if started at time $t$ in the current frame. Here, for the sake of clarity, we are going to consider the symmetric function of $S$: $\hat{S}_i(d) = S_i(D-d)$ gives the frequency for $\task_i$ if this task is started $d$ units of time before the end of the frame. 

In the uniprocessor case, we were able to give schedulability guarantees, as well as good energy consumption performance. We want to be able to provide both in this multiprocessor case, using a global scheduling algorithm. As far as we know, global scheduling algorithm on multiprocessor system using stochastic tasks, and a limited number of available frequencies, has not been considered so far.

The idea of our scheduling algorithm is to consider that a system with $m$ \textcpu, and a frame length $D$, is close to a system with a single \textcpu, but a frame length $m\times D$, or, with a frame length $D$, but $m$ times faster. 
%We show in Figure~\ref{fig:uniproc} what this uniprocessor would look like. 
We then first compute a set of $n$ $\hat{S}$-functions considering the same set of tasks, but a deadline $m\times D$. A very naive approach would consist in considering that when a task ends at time $t$, the total remaining available time before the deadline is the sum of remaining time available on each \textcpu, which means $D-t$ on the current \textcpu, and $D-t_p$ on the other ones, where $t_p$ is the worst time at which the task currently running on $\cpu_p$ will end. Then, we could use $\hat{S}_i(d)$ to choose the frequency. 

Unfortunately, this simple approach does not work, because a single task cannot use time on several \textcpu s \emph{simultaneously}. However, if the number of tasks is reasonably greater than the number of \textcpu s, we think that in most cases, $\hat{S}_i(d)$ will not require to use more than the available time on the current \textcpu, and somehow, will let the available time on other \textcpu s for future tasks. And when $\hat{S}_i(d)$ requires more time than actually available, we just use a faster frequency.

% \begin{figure*}[ht!]
% \caption{Uniprocessor scenario on which the multiprocessor is based.\label{fig:uniproc}}
% \begin{center}
% \includegraphics[width=0.8\linewidth]{uniproc}
% \end{center}
% \end{figure*}

Of course, we need to ensure the schedulability of the system, which cannot be guarantied with the previous approach: for instance, at the end of a frame, we might have some slack time unusable because too short to run any of the remaining task. But as this time has been taken into account when we chose the frequency of previous tasks, we might miss the deadline if we do not take any precaution.

The algorithm we propose is composed of two phases, one off-line, and one on-line. The off-line one consists in performing a (virtual) static partitioning, aiming at reserving enough time in the system for each task. This phase is close to what we did in \cite{RTCSA08} with Danger Zones. The on-line phase uses both this pre-reservation to ensure the schedulability (but performing dynamic changes to this static partitioning), and the $\hat{S}$-functions, to improve the energy efficiency.

\subsection{Virtual Static Partitioning}
We first perform a ``virtual static partitioning''. The aim of this partitioning is not to assign a task to a processor, but to make sure that every task can be executed. A task does not have to run on its assigned processor, but we know that some time has been reserved for this task, which allows to guarantee the schedulability.

This static partitioning can be performed in many ways, but we propose in Algorithm~\ref{alg:statpart} to do it as balanced as possible, by sorting tasks according to their WCEC.

\begin{algorithm}[ht]
\caption{Static partitioning\label{alg:statpart}}\linesnumbered

$A_p = 0 ~\forall p$ ; \tcp{Reserved time on $\cpu_p$} 
$\tasks_p = \{\}~\forall p$ ; \tcp{Tasks assigned to $\cpu_p$}

\ForEach{$\task_i$ descending sorted by $w_i$}{
	$q = \operatorname{argmin}_p A_p$; \tcp{\textcpu{} with the largest not yet assigned time}
	\eIf{$D-A_q > \frac{w_i}{f_M}$}{
		$A_q = A_q + \frac{w_i}{f_M}$; \tcp{$\task_i$ reservation}
		$\tasks_p = \tasks_p \cup \task_i$ \;
	}{
		Failed!
	}
}
\end{algorithm}

\begin{figure}[ht!]
\caption{Left: Static partitioning. Right: State of the system after having started tasks $\{\task_1, \dots, \task_7\}$. Notice that reservations (dashed tasks) correspond to worst cases, while effective tasks (plain lines) are actual execution times, and change then from frame to frame. Vertical axis is frequency, horizontal is time. Then areas correspond to amount of computation. \label{fig:virtpart}}
\begin{center}
\includegraphics[width=0.48\linewidth]{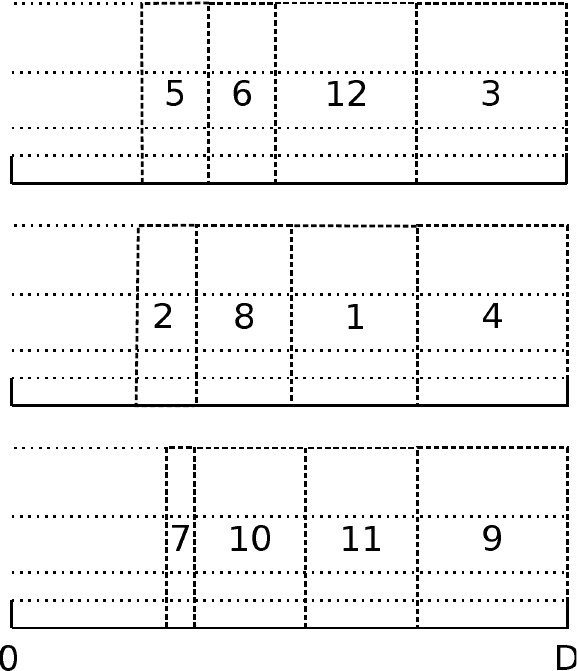} \includegraphics[width=0.48\linewidth]{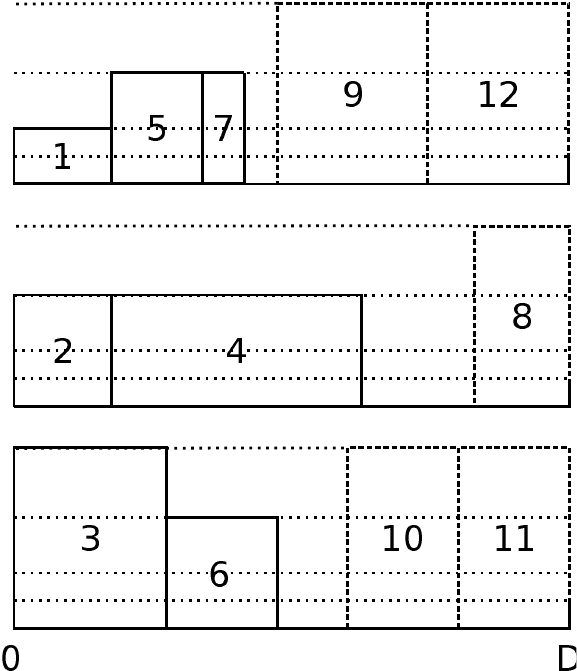}
\end{center}
\end{figure}

After this first step of virtual static partitioning, we can see the system as in Figure~\ref{fig:virtpart}, left part. Notice that it is not because we cannot manage to do this virtual partitioning that the system is not schedulable. But at least, if we manage to do so, then we can ensure that the system is schedulable.
This virtual static partitioning can be computed offline, and used for the whole life of the system.

\subsection{On-line algorithm}

Based on the virtual static partitioning, the main idea of the on-line part is to start a task at a frequency which allows it to end before the beginning of the ``reserved'' part of the frame. For instance, in Figure~\ref{fig:virtpart}, $\task_1$ could start on $\cpu_1$ using all the space between the beginning of the frame, and the reserved space for $\task_5$. But we will see situations where the scheduler needs to give more time for $\task_1$. In such cases, we can also move, for instance, $\task_5$ or  $\task_6$ on $\cpu_2$, or $\task_{12}$ to $\cpu_3$. By doing so, and because we never let a running task using the reserved time of another (not started) task, we can guarantee that, if we were able to build a partitioning in the on-line phase, no task will never miss its deadline. Of course, as soon as a task starts, we release the reserved time for this task.

The on-line part of the algorithm is given in Algorithm~\ref{alg:starttask}. We first give some explanation about two procedures we need in the main algorithm.

\subsubsection{MoveTasksOut}

This procedure (Algorithm~\ref{alg:movetasksout}) aims at moving enough tasks from \textcpu{} $\cpu_p$, until enough space (the quantity $s$ in the algorithm) is available, or no task can be moved anymore. For instance, in Figure~\ref{fig:virtpart}, at time $t=0$, we may want to run $\task_1$ on $\cpu_1$ at frequency $f_2$. But according to the worst case of $\task_1$, we do not have enough time to run this task between 0, and the beginning of the reserved area of $\task_5$. However, we can move $\task_3$ to $\cpu_3$, and $\task_5$ or $\task_6$ to $\cpu_2$. 

While $s$ units of time is not available, we take the largest task on $\cpu_p$, and put it on the \textcpu{} with the largest free space. This is of course a heuristic, since finding the optimal choice is probably NP-hard or at least intractable problem.

\begin{algorithm}[ht]
\caption{MoveTasksOut\label{alg:movetasksout}}\linesnumbered
\KwData{processor $\cpu_p$, current time $t$, space to free $s$}
\tcp{Move out tasks from $\cpu_p$ until $s$ units of time are free from $t$.}

\While{$D-t-A_p \le s$}{
	$\task_i =  $ next task in  $\tasks_p$ (sorted by decreasing $w_i$)\;
	\If{No such $\task_i$}{break\;}

	$q = \operatorname{argmax}_{r\ne p} D-A_r -t_r$; \tcp{\textcpu{} with the maximal amount of available space}
	\If{$D-A_q - t_q > \frac{w_i}{f_m}$}{
		\tcp{Enough place to move $\task_i$ on $\cpu_q$}
		$\tasks_p = \tasks_p \setminus \task_i$ ; $A_p -= \frac{w_i}{f_M}$\;
		$\tasks_q = \tasks_q \cup \task_i$ ; $A_q += \frac{w_i}{f_M}$\;
	}
}
\end{algorithm}

\subsubsection{MoveTaskIn}

This procedure (Algorithm~\ref{alg:movetaskin}) aims at trying to move a task $\task_i$ assigned to some \textcpu{} $\cpu_q$ to the \textcpu{} $\cpu_p$. The main idea is that we first move out as many tasks as needed from $\cpu_p$ (line~\ref{al:mti_mto}), until we have enough space to import $\task_i$ (lines~\ref{al:mti_move} to \ref{al:mti_e_move}). If we have not managed to get enough space, \emph{false} is returned (line~\ref{al:mti_fail}). However, this algorithm is a heuristic, and is not always able to find a solution, even whether such a solution exists. 

For instance (see Figure~\ref{fig:virtpart}, right part), at the end of $\task_7$, we would like to start $\task_8$ on $\cpu_1$. But neither $\task_9$ not $\task_{12}$ can be moved on another \textcpu, so our algorithm fails in finding a solution. However, a smarter algorithm could find out that by swapping $\task_8$ and $\task_9$, $\task_8$ would be able to start on $\cpu_1$. Notice that giving a solution in any solvable case is probably also an NP-hard or at least intractable problem.

The procedure we give here is quite naive, and not very efficient, but we let a better algorithm for further research. The naiveness of this algorithm does not affect the schedulability at all: it just makes the system to be forced more often to accept tasks order changes, which might degrade the energy efficiency ($S$-functions are computed according to the given order), and the user satisfaction, if its preferences are often not respected.

\begin{algorithm}[ht]
\caption{MoveTaskIn\label{alg:movetaskin}}\linesnumbered
\KwData{processor $\cpu_p$, task $\task_i$}
\KwResult{\KwSty{true} if $\task_i$ can be moved on $\cpu_p$, \KwSty{false} otherwise}

\tcp{Move enough tasks from $\cpu_p$ to let $\task_i$ running}
MoveTasksOut($\cpu_p$, $t$, $\frac{w_i}{f_M}$)\; \nllabel{al:mti_mto}

\eIf{$D-t-A_p\ge \frac{w_i}{f_M}$}{\nllabel{al:mti_move}
	let $q$ be such as $\task_i \in \cpu_q$\;
	\tcp{Move $\task_i$ from $\cpu_q$ to $\cpu_p$}
	$\tasks_q = \tasks_q \setminus \task_i$ ; $A_q -= \frac{w_i}{f_M}$\; 
	$\tasks_p = \tasks_p \cup \task_i$ ; $A_p += \frac{w_i}{f_M}$\;
	\Return \KwSty{true}\; \nllabel{al:mti_e_move}
}{
	\Return \KwSty{false}\; \nllabel{al:mti_fail}
}
\end{algorithm}

\subsubsection{Main algorithm}

Here are the main steps of the procedure given in Algorithm~\ref{alg:starttask}, which is called each time a \textcpu{} (say $\cpu_{p}$) is available, at time $t$, with $\task_i$ the next task to start. This procedure will always start at task at a speed guarantying deadlines, but not necessarily $\task_i$.

\begin{itemize}
   \item line \ref{al:d}: We first evaluate $d$, the remaining time we have for $\task_i, \dots, \task_n$: if $t_q$ is the worst time where $\cpu_q$ is going to be available (the time of the last start, plus the worst case execution time of the current task at the chosen frequency), we have:
%    \begin{eqnarray*}
%       d &=& (D-t) + \sum_{q\ne p} (D-t_q) \\
%         &=& P D - \left(t + \sum_{q\ne p} t_q \right) 
%    \end{eqnarray*}
      $$d =  (D-t) + \sum_{q\ne p} (D-t_q) = P D - \left(t + \sum_{q\ne p} t_q \right)\text{.}$$

   \item line \ref{al:f}: Let $f = \hat{S}_i(d)$, the frequency chosen for $\task_i$ in the single \textcpu{} model with $d$ units of time before the deadline. We are going to check if we can use this frequency (we assume this frequency to be a ``good'' one from the energy consumption point of view).
%   \item line \ref{al:e}: Let $e= t+\frac{w_i}{f}$. $e$ is then the (worst) time at which $\task_i$ should end. 
   
   \item line \ref{al:notin}-\ref{al:e_notin}: If $\task_i$ was not assigned to $\cpu_p$, we first try to move it to $\cpu_p$ (Algorithm~\ref{alg:movetaskin}). If we have enough space on $\cpu_p$, the situation is easy. Otherwise, we need to move some tasks out from $\cpu_p$, in order to create enough space.
   
   \item line \ref{al:starti+1}: If we cannot manage to make enough space, then we are not able to start $\task_i$ right now. %This is for instance the case in Figure~\ref{fig:virtpart}, just after the end of $\task_6$. We would like to run $\task_8$, but we do not have enough time before the reservation of tasks $\task_11$ and $\tasks_9$. Furthermore, we cannot move one of those tasks on another CPU, and are then not able to move $\task_8$ on $\cpu_3$.
  We try then the same procedure for $\task_{i+1}$,  but we need to left-shift $\hat{S}-$functions of $\frac{w_i}{f_M}$. This is not required from the schedulability point of view (we ensure the schedulability by controlling the available time), but we guess it will improve the energy consumption. For the same reason, we will need to right-shift functions of the same amount when $\task_i$ starts, because we have one task less to run after $\task_i$. (This improvement is not yet implemented in the given algorithm. It requires to be done carefully, because we might have several swapped tasks).
    
   \item line \ref{al:mto}: If we succeeded, we try to move as many tasks as possible from $\cpu_p$ to other \textcpu{}s (Algorithm~\ref{alg:movetasksout}), until we have enough space to start $\task_i$ at $f$, or no task can be moved anymore. We then start $\task_i$ either at $f$, or at the smallest frequency allowing to run $\task_i$ in the space we manage to free (line~\ref{al:f2}). As $\task_i$ was assigned to $\cpu_p$ (possibly after some changes), we are at least sure that we can start $\task_i$ at $f_M$.
\end{itemize}

Notice that when StartTask is invoked, it is always possible to run a job, and therefore, we will never consider $\tau_{n+1}$ in Algorithm~\ref{alg:starttask}, line~\ref{al:starti+1}. Because of space limitation, we will not give the proof here.
%Here is a sketch of proof. When StartTask is invoked just after the end of a task $\tau_i$ on $\cpu_p$ (not recursively), we have two possibilities: Either some jobs are reserved on $\cpu_p$ between $t$ and $D$, or the interval $[t, D]$ is free. In the first case, at least one of those reserved tasks can start right now. In the second case, if $i\ne n$ (otherwise, we do not have anything to run anymore), the space $[t, D]$ (free) is obviously greater than $[t_r, D]$ for any \textcpu $\cpu_r$ with $r\ne p$. Therefore, we can move to $\cpu_p$ at least one task reserved on another CPU.

% \begin{algorithm}[ht]
% \caption{Using $S$-Functions, Initialisation\label{alg:init}}
% Compute $S$-functions using monoprocessor technics, with $m\times D$ as the frame length\;
% StaticPartitionning() \;
% $t_p=0~\forall p$ \;
% \end{algorithm}

\begin{algorithm}[ht]
\caption{StartTask\label{alg:starttask}}
\restylealgo{boxed}\linesnumbered

%{\small
\KwData{Time $t$, processor $\cpu_p$, task $\task_i$}

$d= P\times D - \left(t + \sum_{q\ne p} t_q \right)$; \tcp{Available time on the system} \nllabel{al:d}
$f = \hat{S}_i(d)$; \tcp{Freq. we want to run $\task_i$} \nllabel{al:f}

\If{$\task_i \notin \tasks_p$}{\nllabel{al:notin}
	\tcp{$\task_i$ is not on $\cpu_p$, we try to move it in}
	\If{\KwSty{not} MoveTaskIn($\cpu_p$, $\task_i$)}{
		StartTask($t$, $\cpu_p$, $\task_{i+1}$)\;\nllabel{al:starti+1}
\Return\;\nllabel{al:e_notin}
	}
}

\tcp{We have now $\task_i \in \tasks_p$}

$A_p -= \frac{w_i}{f_M}$; \tcp{Release $\task_i$ reservation}
$\tasks_p = \tasks_p \setminus \task_i$\;

\tcp{Try to remove enough tasks (if needed) from $\cpu_p$ to allow $\task_i$ to run at the desired speed $f$}
MoveTasksOut($\cpu_p$, $t$, $\frac{w_i}{f}$)\; \nllabel{al:mto}

\If{$D-t-A_p < \frac{w_i}{f}$}{
	\tcp{Not enough time to run $\task_i$ at freq $f$}
	$f = \left\lceil \dfrac{w_i}{D-A_p-t }\right\rceil_\mathcal{F}$\nllabel{al:f2}\;
}

$t_p += \frac{w_i}{f}$; \tcp{Worst end time for $\task_i$}
Start $\task_i$ at $f$\;

%}
\end{algorithm}

\section{Work-in-progress}
Here are a few points we want to look deeper, allowing to improve the energy consumption, or the number of systems we are able to schedule.
\begin{itemize}
   \item At the end of a frame, assuming we can verify that after the task we start, we won't run tasks anymore on this \textcpu{}, we can try to run tasks using the \textcpu{} until $D$. For instance, if we start a task on $\cpu_p$ at a speed which lets a free space $[t_p, D]$ too small to run any of the remaining tasks, then we should try to stretch the task to use $\cpu_p$ up to $D$.
   \item If we accept to change the frequency during the execution of tasks, we can use the continuous model to obtain a frequency $f$, and use two frequencies $\lceil f\rceil_\mathcal{F}$ and $\lfloor f\rfloor_\mathcal{F}$ to ``emulate'' this $f$, where $\lceil f\rceil_\mathcal{F}$ (resp.\ $\lfloor f\rfloor_\mathcal{F}$) stands for the smallest frequency above (resp.\ largest below) $f$.
   \item Several steps require to solve NP-hard problems by using some heuristics: Static partitioning (Algorithm~\ref{alg:statpart}), MoveTaskIn (Algorithm~\ref{alg:movetaskin}), and MoveTasksOut (Algorithm~\ref{alg:movetasksout}). The efficiency of the first one improves the number of systems we can accept to schedule, the second one, the number of tasks we will need to swap (not run in the right order), and the third one, how close we can stay from the uniprocessor algorithm. We may try to improve those three algorithms.
   \item In order to reduce leakage or static energy consumption, we could turn off \textcpu{} if they are not needed anymore before the end of the frame.
\end{itemize}

Of course, we also --- and mainly --- need to validate our model and show its efficiency by the way of simulations, using realistic environment and workloads. 
{\small
\bibliographystyle{acm}
%\bibstyle{plain}
%\nocite{*}

\bibliography{MultiprocFramedDVS}
}

\end{document}